\documentclass[iop]{emulateapj-rtx4} 
\shortauthors{Sekanina}
\shorttitle{Sublimation of Water from Grains in 2I/Borisov?}
\slugcomment{Version November 14, 2019; Addendum November 18, 2019}

\begin{document}
\title{Sublimation of Water Ice from a Population of Large, Long-Lasting
Grains\\Near the Nucleus of 2I/Borisov?}
\author{Zdenek Sekanina}
\affil{Jet Propulsion Laboratory, California Institute of Technology,
  4800 Oak Grove Drive, Pasadena, CA 91109, U.S.A.}
\email{Zdenek.Sekanina@jpl.nasa.gov.}

\begin{abstract}
Potential effects of sublimation of water ice from very slowly moving
millimeter-sized and larger grains. a product of activity at $\sim$10~AU or
farther from the Sun driven presumably by annealing of amorphous water
ice, are investigated by comparing 2I/Borisov with a nominal Oort Cloud
comet of equal perihelion distance of 2~AU.  This comparison suggests
that the strongly hyperbolic motion of 2I mitigates the integrated
sublimation effect.
The population of these grains near the nucleus of 2I is likely to have
been responsible for the comet appearing excessively bright in pre-discovery
images at 5--6~AU from the Sun, when the sublimation rate was exceedingly
low, as well as for the prominent nuclear condensation more recently.
All grains smaller than 2-3~cm across had been devolatilized by
mid-October 2019 and some subjected
to rapid disintegration.  This left only larger chunks of the initial
icy-dust halo contributing to the comet's strongly suspected hyperactivity.
Sublimation of water ice from the nucleus has been increasing
since the time of discovery, but the rate has not been high enough
to exert a measurable nongravitational acceleration on the orbital motion
of 2I/Borisov.
\end{abstract}

\keywords{comets: individual (2I/Borisov, Oort Cloud comets) --- methods: data
 analysis{\vspace{-0.1cm}}}
\section{Introduction}
McKay et al.'s (2019) detection of the red~\mbox{forbidden}
line of atomic oxygen in 2I/Borisov on 2019 October~11
(when the comet was 2.38 AU from the Sun) and interpretation
in terms of the production of water resulted in their
{\vspace{-0.02cm}}determination of a production rate of
0.63\,$\times$\,10$^{27}$\,molecules~s$^{-1}$ with an error
of $\pm$24~percent.  On the other hand, from a 15~hour
integration time of their radio observations of the OH line
at 1667 MHz on October~2--25, Crovisier et al.\ (2019)
{\vspace{-0.05cm}}estimated an OH production rate of 3.3\,$
\times $\,10$^{27}$\,molecules~s$^{-1}$ with an uncertainty
of $\pm$27~percent.  This result is virtually identical with
{\vspace{-0.05cm}}Bolin et al.'s (2019) water production rate
estimate of 100~kg~s$^{-1}$ based on their approximate scaling
of a typical CN/H$_2$O abundance ratio and an available CN
production rate (Fitzsimmons et al.\ 2019).

Combining Crovisier et al.'s production rate result with
Bolin et al.'s estimated upper limit of 1.4~km on the
comet's nuclear diameter and with McKay et al.'s estimate
of an averaged water sublimation rate of 0.37\,$\times
$\,10$^{27}$\,molecules~km$^{-2}$~s$^{-1}$ at the given
heliocentric distance, one finds an active area of more
than 140~percent of the upper limit of the total surface
area of the nucleus.  It appears that the suggestion of the
comet's potential hyperactivity, expressed independently
by McKay et al., is well taken.  In the following,
I present a scenario not considered by McKay et al.\ or
Bolin et al.\ that is in line with a high abundance of
water, without the need of introducing carbon monoxide as
the driver of activity in the pre-discovery observations.

\section{Halo of Large Icy-Dust Grains and Chunks}
In a recent short communication (Sekanina 2019a) I noted that the inner coma
of 2I/Borisov, resolved in the widely disseminated image taken by a camera on
board the Hubble Space Telescope one day after McKay et al.\ (2019) made their
observation, shows a feature extending over a few arcseconds from the nuclear
condensation in a general direction of the negative orbital-velocity vector.
Orbital considerations suggest that this projected side of the inner coma could
be occupied by a halo of nearly-stationary, centimeter-sized and larger icy-dust
grains or pebbles released from the nucleus at times when the comet was still far
(on the order of 10~AU or more) from the Sun.  This type of debris (in a range
from a milli\-meter across up) is commonplace in the characteristic tails of Oort
Cloud comets with perihelia near or beyond the snow line (Sekanina 1975).  The
activity that releases this debris is believed to be driven by annealing of
amorphous water ice in comets just arriving from the Oort Cloud (Meech et al.\
2009).

I now propose that 2I/Borisov has undergone similar evolution as do Oort Cloud
comets on their way to perihelion and that sublimating centimeter-sized and larger
grains and chunks in its atmosphere could contribute substantially to the reported
water production, thereby explaining the comet's hyperactivity even if only a
relatively small fraction of a relatively small nucleus is actually active.  A
very similar scenario was considered by A'Hearn et al.\ (1984) for an Oort
Cloud comet C/1980~E1 (Bowell; old designation 1980~I = 1980b); they attributed
the observed high production rates of water to a cloud of pre-existing sizable
grains proposed in my earlier paper (Sekanina 1982).

To examine the plausibility of applying this hypothesis to 2I/Borisov, I assume
that dark grains, consisting of dirty water ice (i.e., contaminated by refractory
material), are in thermal equilibrium, having a fairly uniform temperature dictated
at a given heliocentric distance by the energy balance between the radiation input
from the Sun and the losses by the thermal reradiation and water ice sublimation.
Somewhat arbitrarily I assume the Bond albedo of 0.04 and the emissivity of 0.9.
Because this sublimation regime differs from the one assumed for the nucleus by
McKay et al.\ (2019), the sublimation area of the putative source of detected
water is also different, equaling 5.8~km$^2$ instead of 1.7~km$^2$.

Let the sublimation rate of water ice implied by the energy balance at time $t$,
{\vspace{-0.05cm}}when the comet is at heliocentric distance $r(t)$, be
$\dot{Z}(t)$, expressed in molecules cm$^{-2}$ s$^{-1}$.  Next, I introduce
{\vspace{-0.05cm}}a columnar sublimation rate, $d\Lambda(t)/dt$ or
$\dot{\Lambda}(t)$, at which the thickness of a layer or the length
{\vspace{-0.02cm}} of a column of water ice contracts (in cm~s$^{-1}$, for
example) due to its sublimation, by{\vspace{-0.05cm}}
\begin{equation}
\dot{\Lambda}(t) = \frac{\mu \dot{Z}(t)}{\delta},{\vspace{-0.1cm}}
\end{equation}
where $\mu$ is the mass of a water molecule and $\delta$ is the bulk density
of water ice in the grain.  Integrating (1) from an early time before the
sublimation began, the length of a column of water ice that has sublimated
away by a reference time, $t_{\rm ref}$, is{\vspace{-0.05cm}}
\begin{equation}
\Lambda(t_{\rm ref}) = \!\! \int_{-\infty}^{t_{\rm ref}} \! \dot{\Lambda}(t) \, dt
 = \frac{\mu}{\delta} \! \int_{\infty}^{r_{\rm ref}} \! \dot{Z}(r) \, \dot{r}^{-1}
 dr ,{\vspace{-0.1cm}}
\end{equation}
where $t_{\rm ref}$ is reckoned from perihelion, $t_\pi$, and $r_{\rm ref}$
is the heliocentric distance at time $t_{\rm ref}$.  In general, there are
no constraints on $t_{\rm ref}$, but in the following I only consider the
contraction of a column of water ice by its sublimation along the preperihelion
leg of the orbit, in which case \mbox{$t_{\rm ref} \!-\! t_\pi \leq 0$} and
\mbox{$\dot{r} \leq 0$}.

To facilitate the integration, I introduce a new, dimensionless variable,
\mbox{$0 \leq z \leq 1$}, by substituting{\vspace{-0.05cm}} 
\begin{equation}
z = \exp \!\left( \frac{\tau}{\tau_0} \! \right),{\vspace{-0.2cm}}
\end{equation}
where \mbox{$\tau = t \!-\! t_\pi$} and \mbox{$\tau_0 > 0$} is a constant.
The length of a column of water ice sublimated by time $t_{\rm ref}$ is then
\begin{equation}
\Lambda(t_{\rm ref}) = \tau_0 \int_{0}^{\exp(\tau_{\rm ref}/\tau_0)} \!
 {\cal F}(t) \, dz,
\end{equation}
where \mbox{$\tau_{\rm ref} = t_{\rm ref} \!-\! t_\pi$} and{\vspace{-0.05cm}}
\begin{equation}
{\cal F}(t) = \frac{\dot{\Lambda}(t)}{z} = \dot{\Lambda}(t) \, \exp \! \left(
 \! -\frac{\tau}{\tau_0} \! \right).
\end{equation}
\begin{figure}[t]
\vspace{-3.95cm}
\hspace{1.55cm}
\centerline{
\scalebox{0.68}{
\includegraphics{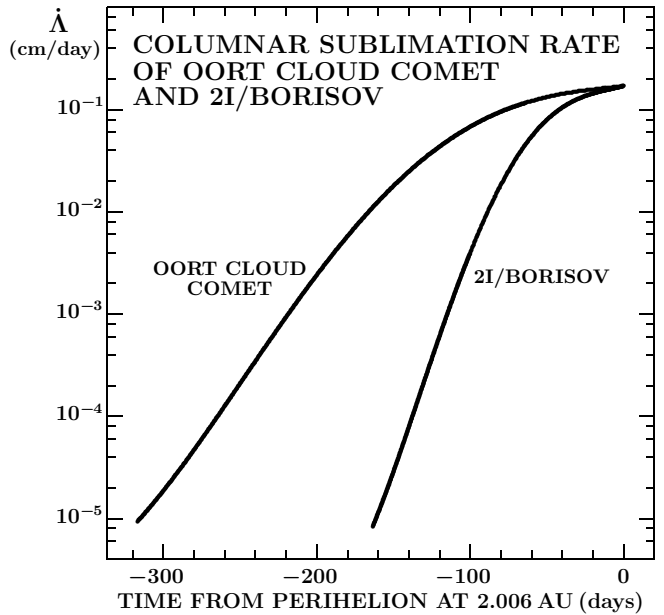}}}
\vspace{-8.4cm}
\caption{Comparison of preperihelion columnar rates of water ice sublimation
from grains in the atmospheres of comet 2I/Borisov and a nominal Oort Cloud
comet of equal perihelion distance of 2.006~AU.  Note that because of the
higher orbital velocity, 2I's rate is lower by one order of magnitude 90 days
before perihelion, but by five orders of magnitude 140~days before
perihelion.{\vspace{0.55cm}}}
\end{figure}

{\vspace{-0.1cm}}
The columnar rates $\dot{\Lambda}(t)$ of water ice sublimation for 2I/Borisov and
an Oort Cloud comet of the perihelion distance of 2.006~AU, are compared in
Figure~1 as a function of time reckoned from perihelion.  The lower rates for
2I/Borisov stem from the higher orbital velocity in its strongly hyperbolic path,
i.e., at an equal time from perihelion 2I is farther from the Sun than the Oort
Cloud comet of equal perihelion distance.

The length of a column of water ice sublimated by the reference time, defined
by Equation~(2), is obviously also shorter for 2I than for the nominal Oort
Cloud comet.  To determine this quantitatively, the integration of the expression
on the right-hand side of Equation~(2) or (4) requires decisions on
two issues:\ (i)~to devise an appropriate method of integration; and (ii)~to
choose the constant $\tau_0$ separately for 2I and the nominal comet in the
parabolic orbit once Equation~(4) is used.

The choice of $\tau_0$ is not at all critical, the only concern being that, for
{\vspace{-0.05cm}}the purpose of smooth integration, ${\cal F}$ is approximately
flat near perihelion, where $\dot{\Lambda}$ reaches its maximum.  This condition
is satisfied by \mbox{$\tau_0 = 150$ days} for 2I/Borisov and by \mbox{$\tau_0
= 300$ days} for the nominal Oort Cloud comet in the parabolic orbit.

An obvious choice for approximating a function defined by a sequence of
\mbox{($x$, $y$)} data pairs is a polynomial $y(x)$, whose coefficients are
determined by least squares.  The advantage of this approach is twofold:\
first, one can optimize the polynomial's degree by searching for the lowest
possible degree that secures the requested quality of fit; and second, the
integration is straightforward.  To follow this procedure, I write\\[-0.25cm]
\begin{equation}
{\cal F}_j(t) = \sum_{k=0}^{n} a_{k,j} z^k,\\[-0.1cm]
\end{equation}
where \mbox{$j = 1$} for 2I/Borisov and \mbox{$j = 2$} for the Oort Cloud comet.
In accordance with this notation, I refer from now on to $\Lambda$ and $\tau_0$
for 2I/Borisov as $\Lambda_1$ and $\tau_{0,1}$, for the Oort Cloud comet as
$\Lambda_2$ and $\tau_{0.2}$, respectively.  Experimentation with a number of
reference points ($z$, ${\cal F}$) has shown that a cubic polynomial fits
both 2I/Borisov and the Oort Cloud comet quite adequately,
if a typical error of approximately $\pm$0.001 cm~day$^{-1}$ for ${\cal F}_j$
is acceptable, but {\it only\/} over the range of heliocentric distances not
exceeding \mbox{$r^{\displaystyle \ast} = 3.4$~AU},
equivalent to about \mbox{$z_1^{\large \ast} \simeq 0.43$}
and \mbox{$t_1^{\displaystyle \ast} \!-\! t_\pi = -125$ days} for
{\vspace{-0.04cm}}2I/Borisov and to \mbox{$z_2^{\large \ast} \simeq 0.45$} and
\mbox{$t_2^{\displaystyle \ast} \!-\! t_\pi = -240$ days} for the nominal Oort
Cloud comet.  The failure of any fit of this kind for all values of $z$ smaller
(and times from perihelion larger) than these limits has two major implications.
One is the need to devise an integration approach other than via Equation~(4)
{\vspace{-0.05cm}}for all reference times preceding $t^{\displaystyle \ast}$
(i.e., for {\vspace{-0.05cm}}heliocentric distances larger than
$r^{\displaystyle \ast}$).  And two, all reference times between
$t^{\displaystyle \ast}$ and the perihelion time $t_\pi$
call obviously for a revision of the integration limits in Equation~(4).

The remedy of the first problem is greatly facilitated by the fact that at
heliocentric distances exceeding 3.4~AU the incident solar radiation is spent
overwhelmingly on reradiation, so that the columnar rate of water ice
sublimation follows closely a relation{\vspace{-0.1cm}}
\begin{equation}
\dot{\Lambda} = A \exp \!\left( \! -B \sqrt{r} \,\right),
\end{equation}

{\vspace{-0.2cm}} \noindent
where \mbox{$A = 6.7 \times \! 10^{13}$ cm day$^{-1}$} and \mbox{$B = 21.587$
AU$^{-\frac{1}{2}}$}.  One bypasses Equation~(4) by integrating directly
Equation~(2) to obtain for the length of the sublimated{\vspace{-0.04cm}}
column of water ice by time $t^{\displaystyle \ast\!}$ an expression
\begin{equation}
\Lambda_j(t^{\displaystyle \ast}) = A \!\! \int_{-\infty}^{t^{\scriptstyle \ast}}
 \!\!\! \exp \! \left( \! -B \sqrt{r} \, \right) dt .
\end{equation}
Next one substitutes \mbox{$dt = \dot{r}^{-1} dr$} with
\begin{equation}
\dot{r} = \frac{k_0 e \sin u}{\sqrt{p}},
\end{equation}
where \mbox{$k_0 = 0.0172021$ AU$^{\frac{3}{2}}$ day$^{-1}$} is the Gaussian
gravitational constant, $e$ the orbit eccentricity, \mbox{$p = q \:\!(1 \!+\! e)$}
the semi-latus rectum (in AU), $q$ the perihelion distance (in AU), and $u$
{\vspace{-0.02cm}}the true anomaly.  At a heliocentric distance of 3.4~AU
preperihelion one obtains \mbox{$\dot{r} = -29.88$ km s$^{-1}$} for 2I/Borisov,
{\vspace{-0.04cm}}but $-$14.63~km~s$^{-1}$ for the nominal Oort Cloud comet.
Examining $\Lambda_2(t_2^{\displaystyle \ast})$ first, I note that for a parabolic
orbit Equation~(9) becomes before perihelion
\begin{equation}
\dot{r} = -\frac{k_0 \sqrt{2}}{r} \sqrt{r \!-\! q},
\end{equation}
which for \mbox{$r > fq$} satisfies a condition
\begin{equation}
|\dot{r}|^{-1} < \frac{\sqrt{r}}{k_0 \sqrt{2}} \sqrt{\frac{f}{f \!-\! 1}}.
\end{equation}
Inserting Equation (11) into Equation (8) I find
\begin{eqnarray}
\Lambda_2(t_2^{\displaystyle \ast}) & < & \frac{A}{k_0 \sqrt{2}} \, \sqrt{\frac{f}
 {f \!-\! 1}} \int_{r^{\scriptstyle \ast}}^{\infty} \!\! \sqrt{r} \exp \! \left(
 -B \sqrt{r} \, \right) dr \nonumber \\[-0.2cm]
 & & \\[-0.2cm]
 & = & \frac{2A \sqrt{2}}{k_0 B^3} \sqrt{\frac{f}{f
 \!-\! 1}} \exp \! \left( \!-B \sqrt{r^{\displaystyle \ast}} \right)
 \!\! \left( \! 1 \!+\! B \sqrt{r^{\displaystyle \ast}} \!+\! {\textstyle
 \frac{1}{2}} B^2 r^{\displaystyle \ast} \! \right)\! . \nonumber
\end{eqnarray}
For \mbox{$f = r^{\displaystyle \ast\!\!}/q = 1.695$} I find
\begin{equation}
\sqrt{\frac{f}{f \!-\! 1}} = \sqrt{\frac{r^{\displaystyle \ast}}
{r^{\displaystyle \ast\!} \!-\! q}} = 1.562.
\end{equation}
Inserting the numerical values into Equation~(12),~one obtains
\mbox{$\Lambda_2(t_2^{\displaystyle \ast}) < 0.0073$ cm} for the nominal Oort
Cloud comet and from the hyperbolic-to-parabolic ratio of the radial
velocities \mbox{$\Lambda_1(t_1^{\displaystyle \ast}) < 0.0036$ cm} for
2I/Borisov, both implying negligible sublimation losses along either
trajectory for centimeter-sized grains by the time the object had reached 3.4~AU.

The second remedy, required by the limits of the polynomial approximation (6) to
the columnar sublimation rate, is predictably a relatively minor modification
of Equation~(4).  Marking now with primes the lengths of the sublimated columns
computed by integrating ${\cal F}_j$ via polynomials, $\Lambda_j^\prime$,
I find (dropping the subscript {\it ref\/})
\begin{eqnarray}
\Lambda_j(t) & = & \tau_{0,j} \!\!\int_{\exp(\tau_j^{\scriptstyle
 \ast\!}/\tau_0)}^{\exp(\tau_j/\tau_0)} \!\! {\cal F}_j \, dz +
 \Lambda_j(t_j^{\displaystyle \ast}) \nonumber \\[-0.25cm]
 & & \\[-0.15cm]
 & = & \tau_{0,j} \!\left[ \Lambda_j^\prime(t)\!-\!
 \Lambda_j^\prime(t_j^{\displaystyle \ast\!}) ] + o[\Lambda_j(t)
 \right], \;\;\; j = 1, 2, \nonumber
\end{eqnarray}
{\vspace{-0.05cm}}where \mbox{$\tau_j^{\displaystyle \ast} = t_j^{\displaystyle
\ast} \!-\! t_\pi$}, $o[x]$ means a negligible contribution to $x$, and again
\mbox{$j = 1$} for 2I/Borisov and \mbox{$j = 2$} for the nominal Oort
Cloud comet.

I already remarked that cubic polynomials represent adequate approximations
for both objects; the columns $\Lambda_j^\prime$ sublimated by time $t$
(\mbox{$t_j^{\displaystyle \ast} < t \leq t_\pi$}),{\vspace{-0.08cm}}
employed in Equation~(14), equal:
\begin{equation}
\Lambda_j^\prime(t) = \tau_{0,j} z \sum_{k=0}^{3} \frac{b_{k,j} z^{k}}{k+1},
 \;\;\; j = 1, 2,
\end{equation}
where
\begin{equation}
b_{k,j} = \frac{a_{k,j}}{k+1}, \;\;\; k\!=\!0, \ldots 3; \;\; j \!=\! 1, 2,
\end{equation}
and
\begin{eqnarray}
        &   &   \;\;\;\; j = 1 \,\;\;\;\;\;\;\;\;\;\; j = 2 \nonumber \\
b_{0,j} & = & +0.78485, \;\;\; +0.96249, \nonumber \\
b_{1,j} & = & -2.06856, \;\;\; -2.41993, \nonumber \\
b_{2,j} & = & +2.26169, \;\;\; +2.53858, \nonumber \\
b_{3,j} & = & -0.81706, \;\;\; -0.89328.
\end{eqnarray}
\begin{table}[b]
\vspace{-3.8cm}
\hspace{4.2cm}
\centerline{
\scalebox{1}{
\includegraphics{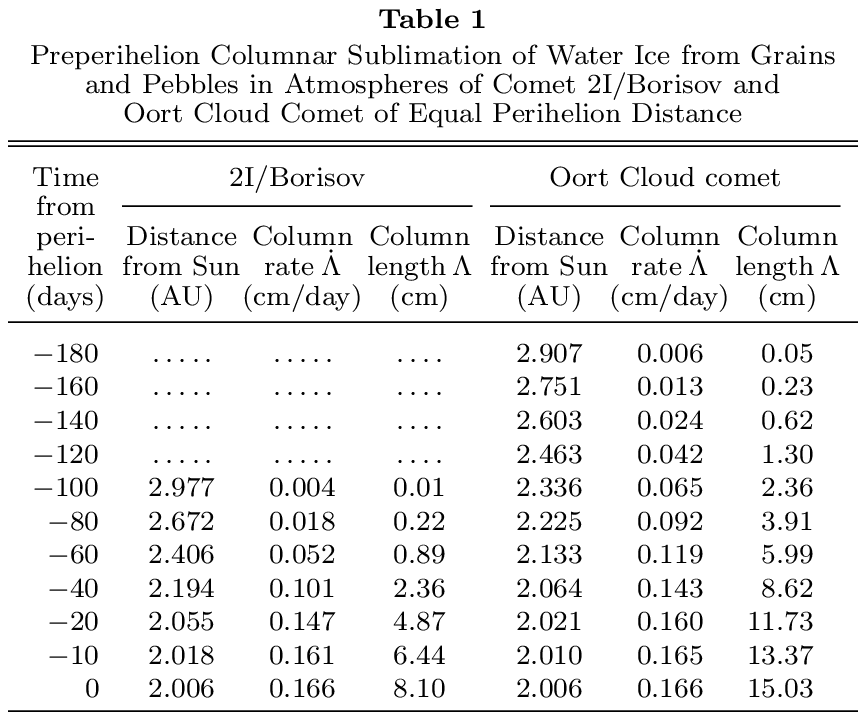}}}
\vspace{-18.5cm}
\end{table}

Table 1 compares 2I/Borisov with the nominal Oort Cloud comet in terms of both
the columnar rate of water ice sublimation and the total length of sublimated
column of ice at various times before perihelion.  Figures 2 and 3 show the
sublimated columnar lengths as a function of time and heliocentric distance,
respectively.
\begin{figure}[t]
\vspace{-2.25cm}
\hspace{0.65cm}
\centerline{
\scalebox{0.66}{
\includegraphics{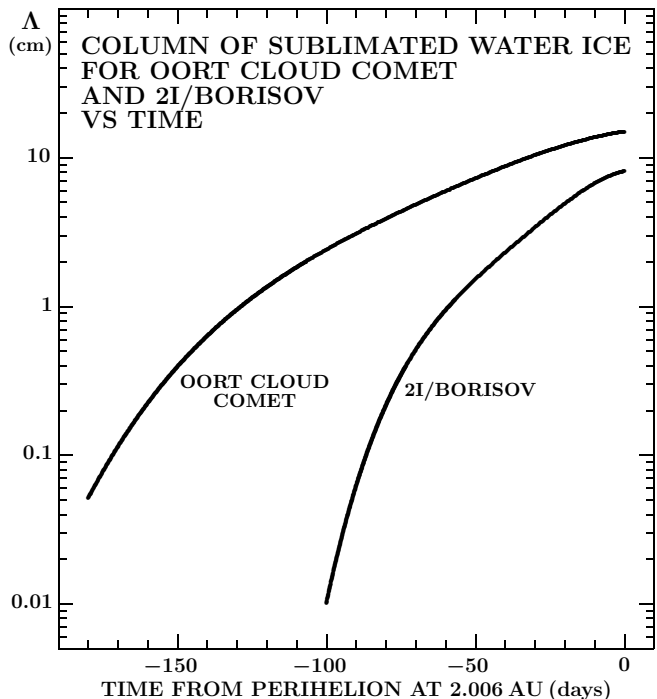}}}
\vspace{-8.37cm}
\caption{Column of water ice sublimated away from a chunk as
a function of time before perihelion for 2I/Borisov and an Oort Cloud comet
of equal perihelion distance.{\vspace{0.5cm}}}
\end{figure}
Large differences between the two objects are plainly apparent, especially
when plotted against time.  For example, by the time of the Hubble Space
Telescope's observation, 57 days before perihelion, the column of water ice
that sublimated away should have been about 1~cm, so that ice would have
been gone from all grains of up to about 2~cm across.  According to Table~1
of Sekanina~(2019a),~such chunks would be at 6300~km from the nucleus if they
had been released $\sim$1~year before perihelion, at $\sim$8~AU from
the Sun.  By contrast, an Oort Cloud comet of equal perihelion distance would
have lost by that time a column of water ice more than 6~cm thick, so only very
massive chunks, much larger than 10~cm across would still possess some remaining
ice.  Thus, the substantially higher orbital velocity and the hyperbolic shape
of the trajectory have helped 2I/Borisov lose water ice at a significantly
lower rate than does an Oort Cloud of equal perihelion distance.{\vspace{0.05cm}}

\begin{figure}[t]
\vspace{-2.25cm}
\hspace{0.65cm}
\centerline{
\scalebox{0.66}{
\includegraphics{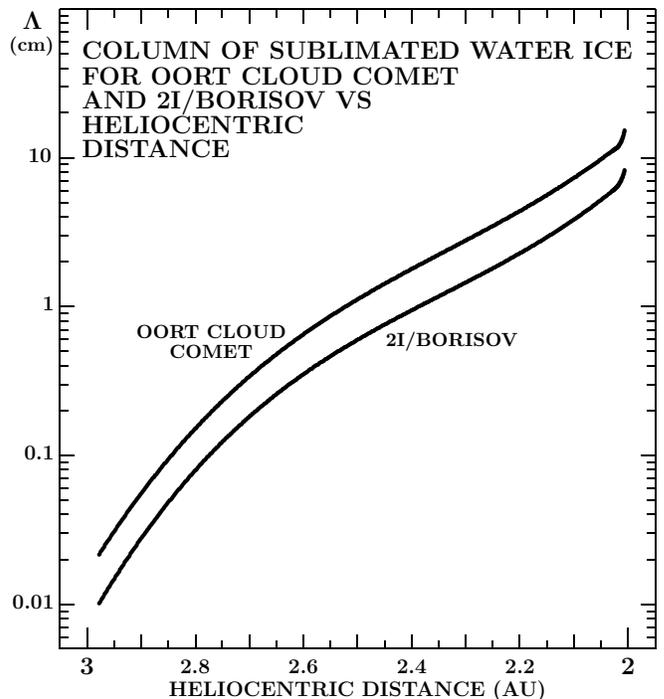}}}
\vspace{-8.37cm}
\caption{Column of water ice sublimated away from a chunk as
a function of heliocentric distance before perihelion for 2I/Borisov and an
Oort Cloud comet of equal perihelion distance.{\vspace{0.6cm}}}
\end{figure}

\section{Population of Large Grains and the Hyperactivity}
The term {\it hyperactive comets\/} is usually employed to
refer to a small group of short-period comets, including
21P/Giacobini-Zinner, 45P/Honda-Mrkos-Pajdu\v{s}\'akov\'a,
46P/Wirtanen, and 103P/Hartley, whose water production is
anomalously high, often exceeding the level that implies the
entire surface area of the nucleus is active.  The best examined
member of this group is 103P, the target of the EPOXI
mission, whose hyperactivity was proposed by A'Hearn et al.\
(2011) to be driven primarily by gaseous CO$_2$ that blasts
off chunks of sublimating water ice.  Yet, Harker et al.\
(2018) have concluded that the underlying cause of the
hyperactivity is still unknown.

Independently of the outcome of this dispute, I believe that 
103P or any other short-period comet is not an appropriate
standard for investigating the hyperactivity of 2I/Borisov
because of their very different histories.  A much better
benchmark is C/1980~E1 (Bowell), an Oort Cloud comet with
perihelion at 3.36~AU, which turned out to be extremely
hyperactive around 5~AU preperihelion, although this
term was not~yet~in~use~in~those~days.  A'Hearn et al.\
{\vspace{-0.05cm}}(1984) reported in their paper a water
production rate of 0.74\,$\times$\,10$^{29}$~molecules~s$^{-1}$
{\vspace{-0.01cm}}at 5.25~AU
and 1.55\,$\times$\,10$ ^{29}$~molecules~s$^{-1}$ at 4.63~AU
from the Sun.  These rates imply the surface area of a
nonrotating nucleus of, respectively, 460~km and 190~km in
diameter (!) and even a larger rapidly rotating nucleus.
Although the nuclear size of C/1980~E1 is unknown, it has
been estimated at not more than 10~km by A'Hearn et al.\
(1984), a discrepancy of more than one order of magnitude.

A'Hearn et al.\ resolved this disparity by assuming that
the production of water near 5~AU preperihelion came from
a source other than the nucleus, referring to my work on a
population of large grains ($>$0.5~mm across) in the coma
and tail of C/1980~E1 released from the nucleus probably near
\mbox{11--12}~AU from the Sun (Sekanina 1982), a product of
the comet's activity on its first journey to the inner Solar
System.  Estimates of the total mass of this population of
grains ranged from more than 10$^{13}$~g to 5\,$\times$\,10$
^{15}$~g, corresponding to a layer of between $\sim$1~cm and
$\sim$1~m on a 10~km nucleus.

Grains observed in the tail of an Oort Cloud comet, whose
perihelion is near or beyond the snow line, are released from
the nucleus over a period of time, but the tail's narrow width
and the characteristic major gap between the tail and the antisolar
direction demonstrate that the grain release stopped long
before observation.  The time it began is unfortunately not
well determined because of poor temporal resolution among
early emissions in the tail orientation.  Yet, I showed on
an example of comet C/1954~O1 (Baade) that measuring the
position angle of the tail's axis (as is commonly done)
instead of the position angle of its maximum extent has a
tendency to significantly underestimate both the minimum
grain size in the tail and the heliocentric distance at
release (Sekanina 1975).  For any comet displaying such a
tail before perihelion, the heliocentric distance at release
must obviously exceed markedly the perihelion distance.
While Oort Cloud comets discovered \mbox{60--70} years ago
had perihelia at distances of up to at most 5~AU, more
recently the limit has moved up by several~AU, yet some of
these objects still display the tails, some even before
perihelion.\footnote{An excellent example of what can be
learned from high-quality imaging obtained at an appropriate
time is Meech et al.'s (2009) image of C/1999~J2, taken on
2000 February 24, or 42~days pre\-perihelion, at 7.11~AU.
Compared to the other five images of this comet in the
paper, the great advantage of this one is the large angle
between the radius vector (p.a.\ 279$^\circ$) and the
negative orbital velocity vector (p.a.\ 20$^\circ$).  For
the left, sharper, and perfectly rectilinear boundary of
the tail I measure a position angle of 13$^\circ$, which
implies a release time of 1600~days before perihelion at
12~AU from the Sun.  Assuming the tail extends to the edge
of the frame, the smallest grains are subjected to a
radiation pressure acceleration of 0.0014 the Sun's
gravitational acceleration and are about 1.6~mm across.
Their temperature is about 80\,K.  The right, much shorter,
and less sharp boundary of the tail extends to a position
angle of about 2--3$^\circ$, implying a terminal release
time 600~days before perihelion and a heliocentric distance
of $\sim$8~AU.  The grains'~implied temperature is about 100\,K.
The process extended for approximately 1000 days with an
activating temperature range~of~$\sim$20$^\circ$.} This is
a very strong argument for the onset of release of grains
from Oort Cloud comets at heliocentric distances of at least
$\sim$10~AU from the Sun, corresponding to temperatures of
not more than 80\,K for a rapidly rotating comet and lower
than 100\,K for a nonrotating comet.

A paper by Meech et al.\ (2009) addresses the problem of
the potential mechanism of large-grain release at very low
temperatures; they prefer
annealing of amorphous water ice, which is stable up to at
least 120--130\,K, over sublimation of carbon monoxide as
the driver of activity far from the Sun in Oort Cloud comets.
They show, in fact, that in laboratory experiments the
annealing process begins at a temperature as low as 37\,K,
corresponding to a heliocentric distance of 59~AU for a
rapidly rotating (isothermal) nucleus and still farther from
the Sun for a nonrotating nucleus.  Ice physics thus provides
only modest constraints in terms of heliocentric distance at
the time of release of grains.

Additional issues relate to the grains.  Relative fractions
of water ice, other ices, and refractory material in the
grains upon arrival from the Oort Cloud are unknown.  Nor
is  the grains' degree of coherence and their fate after the
evacuation of water ice:\ does the skeleton hold together or
does it disintegrate?  And if it holds together, how does
the loss of water ice affect its optical properties, such as
the albedo?  On the last issue, there is evidence that might
indicate either the grains' failure to survive without the
ice or a major drop in their reflectivity because the tails
have a tendency to disappear in comets with perihelion
distances below the snow line, being replaced near and after
perihelion with tails of lesser age.  Yet, these tails also
lack grains substantially smaller than $\sim$1~mm and no
post-perihelion emission of dust is typically detected.

\section{Population of Large Grains and Chunks in 2I/Borisov and Absence
 of Nongravitational Acceleration in Its Orbital Motion}
Bolin et al.\ (2019) reported Ye et al.'s (2019) detection of
pre-discovery $r$-filter images of 2I/Borisov on 2019 March
17--18 and May 2 and 5 (when the comet was between 6.03~AU
and 5.09~AU from the Sun), taken with the Zwicky Transient
Facility's (ZTF) wide-field camera mounted on the 122-cm
Schmidt telescope at Palomar, thus extending the observed arc
of the orbit by 5.5~months.  The authors pointed out that the
comet was at the time of magnitude 20.5 to 21.0, much brighter
than predicted by a water ice sublimation model, and argued that
the brightness was consistent with activity driven by sublimation
of carbon monoxide.  The plot in their Figure~6 shows, however, that
between 6~AU and 2.5~AU from the Sun the {\vspace{-0.07cm}}scattering
cross-sectional area then varied with heliocentric distance $r$ as
$r^{-\frac{2}{3}}$~rather~than~the expected $\sim \! r^{-2}$ or
steeper.  And because activity at 2.5~AU was
driven by sublimation of water ice and other volatiles from the
nucleus, the exponent for the CO model should be still closer to
zero.  In addition, missing in Bolin et al.'s Figure~6 is evidence
for a definite increase in the scattering cross section of
dust from March to May, which would be expected in the case
of CO driven activity.

On the other hand, if the comet's excessive brightness observed
in the ZTF images of 2I/Borisov was due to a population of
sizable, extremely slowly moving grains, the product of activity
at very large heliocentric distance driven presumably by
annealing of amorphous water ice (Meech et al.\ 2009), one
would expect the cross-sectional area to stay constant between
2019 March and May.  For an assumed albedo of 0.04, Bolin et
al.'s results suggest the total cross-sectional area of the
comet to equal $\sim$200~km$^2$ at \mbox{5--6}~AU before
perihelion.  Assuming this refers to the grain halo and a
typical grain diameter of $\sim$2~mm, the
volume is 3\,$\times$\,10$^{11}$\,cm$^{3}$, which makes a
layer of more than 10~cm thick on the nucleus smaller than
1.4~km across, in the same range as the result for
the population of grains in C/1980~E1 (Section 3).

Because the sublimation rate of water ice from grains in the
atmosphere of 2I/Borisov increases exponentially with time
at heliocentric distances exceeding $\sim$3~AU, reaching an
integrated columnar length of 0.1~cm near 2.77~AU, 1~cm near
2.39~AU, 2~cm near 2.23~AU, and 4~cm near 2.1~AU (Figure~3),
grains of the respective sizes become devolatilized at the
respective heliocentric distances and subject to potential
disintegration.  This could explain three additional points:\
(i)~the presumed presence, in the Hubble Space Telescope's
image taken on October~12, of a population of centimeter-sized
and larger chunks in the inner coma (Sekanina 2019a) and,
simultaneously, the absence of a tail-like extension,
composed of millimeter-sized grains of the same population and
pointing along the direction of the negative orbital-velocity
vector; (ii)~the comet's hyperactivity over a broad range of
heliocentric distances around 2.5~AU preperihelion, with a major
contribution by the same centimeter-sized and larger grains
because of their water sublimation; and (iii)~the bumps in the
comet's light curve in the course of September and early October
(Bolin et al.\ 2019), which could be triggered by clouds of
microscopic-sized  debris of large devolatilized grains
disintegrating at temporally variable rates. 

Bolin et al.\ (2019) appear to admit the presence of a
nongravitational acceleration in the orbital motion of 2I/Borisov
by noting that ``[m]oderate non-gravitational force parameters
have been measured for the orbit of 2I in pre-discovery data when
the comet's activity was weaker (Ye et al.\ 2019).''  However,
consideration of the reported outgassing rates of 2I suggests that
after normalization to 1~AU from the Sun, the nongravitational
acceleration on this comet should at best be two orders of magnitude
lower than the observed anomalous effect on 1I/`Oumuamua.  In fact,
Nakano (2019) linked the pre-discovery astrometry from the ZTF database
for 2018 December~13 and 2019 February~24, March~17, April~9,
and May~2 and 5 with more than 1600~observations from August~30
through November~6 with no need to incorporate nongravitational
terms into the equations of motion.  His purely gravitational
solution provides an excellent fit to the dataset with a mean
residual of $\pm$0$^{\prime\prime\!}$.67 and a very satisfactory
distribution of individual residuals (which are explicitly
tabulated by Nakano), free from any systematic trends.

\section{Conclusions}
While I by no means rule out sublimation of~\mbox{carbon} monoxide (or
other volatiles not directly~tied~to~sublimation of water ice)
from 2I/Borisov, I see no compelling evidence from the pre-discovery
observations for CO driving the comet's activity before water ice
sublimation began to gradually dominate, as documented by the data
acquired since discovery.  I find that the
observations reported thus far are consistent with a pre-existing
population of (initially) millimeter-sized and larger icy-dust
grains and chunks in the coma, presumably a product of the
process of annealing of amorphous water ice (Meech et al.\ 2009)
that proceeded in the conglomerate nucleus when the comet was, say,
10~AU or farther from the Sun.  The presence of such a massive
grain population made the comet bright in the 2019 March--May
images, at 5--6~AU from the Sun.  Devolatilization and subsequent
disintegration of the smallest, millimeter-sized grains from this
population was being completed in the course of September of 2019,
and their microscopic-sized relics were gradually but temporally
unevenly eliminated from the comet's proximity by solar
radiation pressure.

By mid-October, eight weeks before perihelion, sublimation of water
continued only from the halo's~chunks $>$2--3~cm in diameter, the
grain population's sublimation cross section dropping from the
initial $\sim$200~km$^2$ down to $<$6~km$^2$ or $<$30~km$^2$,
depending on whether one uses McKay et al.'s (2019) or Crovisier
et al.'s (2019 data.  Before the comet reaches perihelion, water
sublimation from the nucleus should begin to dominate and the
hyperactivity vanish, if 2I evolves as do Oort Cloud comets.
As of mid-November 2019, the activity of 2I/Borisov has closely
replicated --- except for the reduced sublimation effect --- that
of Oort Cloud comets of similar perihelion distance, as previously
suspected (Sekanina 2019b); it remains to be seen whether this
affinity is going to be maintained throughout~2I's~journey
about the Sun.\\[-0.1cm]

This research was carried out at the Jet Propulsion Laboratory, California
Institute of Technology, under contract with the National Aeronautics and
Space~Administration. \\[-0.1cm]

\begin{center}
{\footnotesize REFERENCES}
\end{center}
\vspace{-0.15cm}
\begin{description}
{\footnotesize
\item[\hspace{-0.3cm}]
A'Hearn, M.\ F., Schleicher, D.\ G., Feldman, P.\ D., et al.\ 1984, AJ,{\linebreak}
 {\hspace*{-0.6cm}}89, 579
\\[-0.57cm]
\item[\hspace{-0.3cm}]
A'Hearn, M.\ F., Belton, M.\ J.\ S., Delamere, W.\ A., et al.\ 2011,{\vspace{-0.06cm}\linebreak}
 {\hspace*{-0.6cm}}Science, 332, 1396 (erratum, 2012)}
\end{description}
\begin{description}
{\footnotesize
\item[\hspace{-0.3cm}]
Bolin, B.\ T., Lisse, C.\ M., Kasliwal, M.\ M., et al.\ 2019, eprint{\linebreak}
 {\hspace*{-0.6cm}}arXiv:1910.14004
\\[-0.57cm]
\item[\hspace{-0.3cm}]
Crovisier, J., Colom, P., Biver, N., et al.\ 2019, CBET 4691
\\[-0.57cm]
\item[\hspace{-0.3cm}]
Fitzsimmons, A., Hainaut, O., Meech, K.\ J., et al.\ 2019, ApJ, 885,{\linebreak}
 {\hspace*{-0.6cm}}L9
\\[-0.57cm]
%
%
\item[\hspace{-0.3cm}]
Harker, D.\ E., Woodward, C.\ E., Kelley, M.\ S.\ P., et al.\ 2018, AJ,{\linebreak}
 {\hspace*{-0.6cm}}155, 199
\\[-0.57cm]
\item[\hspace{-0.3cm}]
McKay, A.\ J., Cochran, A.\ L., Dello Russo, N., et al.\ 2019, eprint{\linebreak}
 {\hspace*{-0.6cm}}arXiv:1910.12785
\\[-0.57cm]
\item[\hspace{-0.3cm}]
Meech, K.\ J., Pittichov\'a, J., Bar-Nun, A., et al.\ 2009, Icarus, 201,{\linebreak}
 {\hspace*{-0.6cm}}719
\\[-0.57cm]
\item[\hspace{-0.3cm}]
Nakano, S.\ 2019, NK 3929
\\[-0.57cm]
\item[\hspace{-0.3cm}]
Sekanina, Z.\ 1975, Icarus, 25, 218
\\[-0.57cm]
\item[\hspace{-0.3cm}]
Sekanina, Z.\ 1982, AJ, 87, 161
\\[-0.57cm] 
\item[\hspace{-0.3cm}]
Sekanina, Z.\ 2019a, eprint arXiv:1910.11457
\\[-0.57cm] 
\item[\hspace{-0.3cm}]
Sekanina, Z.\ 2019b, eprint arXiv:1910.08208
\\[-0.63cm]
\item[\hspace{-0.3cm}]
Ye, Q.-Z., Kelley, M.\ S.\ P., Bolin, B.\ T., et al.\ 2019, in preparation}
\vspace{-0.2cm}
\end{description}
\vspace{0cm}
\begin{center}
{\bf ADDENDUM, dated November 18, 2019}
\end{center}
{\hspace{0.28cm}}
Now that the results of the pre-discovery observations investigated by
Ye et al.\ (2019, eprint arXiv:1911.05902) have been published, two of
the highlights of their work stand out as the most diagnostic:\ (i)~a
ballooning of the cross-sectional area of the comet from an undetected
level (with a $3\sigma$ upper limit of 140~km$^2$) in{\vspace{-0.04cm}}
November 2018 to about 330~km$^2$ a month later at heliocentric distances
$r$ near 8~AU; and (ii)~the comet's intrinsic{\vspace{-0.05cm}}
brightness, which varies approximately as $r^{-2}$ from 8~AU down all
the way to 2.4~AU, implies an essentially constant cross-sectional area
over this range of heliocentric distances, until at least early October
2019, a 10-month period.

The most straightforward interpretation of the findings --- which should
likewise accommodate the high production rate of water in October --- is
liberation, over a period of several weeks to a few months and at
extremely low velocities, of a large amount of icy-dust debris from the
nucleus into the atmosphere, where it has been lingering for the 10 months
seemingly unaltered.  The process of annealing as a driver of this activity,
proposed by Meech et al.\ (2009), was based on Bar-Nun et al.'s laboratory
experiments, but independent and/or more recent work on annealing
[e.g., S.\ A.\ Sandford \& L.\ J.\ Allamandola {\vspace{-0.06cm}}(1988,
Icarus, 76, 201), B.\ Schmitt et al.\ (1989, ESA SP-302, 65), O.'
\'O.\ G\'alvez et al.\ (2008, Icarus, 197, 599), L.\ J.\ Karssemeijer et al.\
(2013, ApJ, 781, 16), R.\ Mart\'{\i}n-Dom\'enech et al.\ (2014, A\&A, 564,
A8), A.\ N.\ Greenberg et al.\ (2017, MNRAS, 469, S517)] does in general
support this suggestion.  It is possible that the process was already in
progress in November 2018 or perhaps even earlier, lasting for a few months.
Some of the experiments indicate that especially CO$_2$/H$_2$O ice mixtures
exhibit a strong annealing effect above 90\,K, equivalent to the relevant
range of heliocentric distances beyond 8~AU.  Crystallization of amorphous
ice, which also could trigger activity, should start near the subsolar point
at about the same time or soon afterward.  One has to keep in mind a
potentially large temperature variations over the surface of the nucleus.
 
The requirement of extremely low velocities implies large size of the
released debris.  An order-of-magnitude estimate for the velocities is
derived from an extreme condition that after 10~months the chunks should
stay confined to within a fairly small distance, say, 5000 to 10\,000~km,
of the nucleus, even if they keep expanding; this
suggests 0.2 to 0.4~m~s$^{-1}$, comparable to the velocity of escape from
a small comet.  If an average piece of debris is between 0.1~cm
and 1~cm across and of a bulk density of 0.5~g~cm$^{-3}$, the estimated
mass of fragments with the cross-sectional area of 330~km$^2$ is
(\mbox{1--10})\,$\times$10$^{11}$\,g.  On a nucleus of {\it less\/}
than 1.4~km across (Bolin et al.\ 2019) and equal density, this mass would
be distributed in a layer of {\it more\/} than \mbox{3--30 cm} thick, an
estimate whose upper bound is in order-of-magnitude agreement with the
thickness of a layer of grains released from the nucleus of C/1980~E1,
estimated by A'Hearn et al.\ (1984) from their determination of the comet's
sublimation rate of water ice near 5~AU preperihelion.

The pre-discovery observations refer to a total {\it scattering\/} cross
section of the debris, offering no information of the {\it sublimation\/}
cross section, which is assumed to be the same near 8~AU, but dropping with
decreasing heliocentric distance because of progressive depletion of water
ice from smaller fragments by increasing sublimation nearer perihelion, as
shown in Table~1.  It is still unclear whether, or at what rate, does the
scattering cross section of the devolatilized debris drop with time.
Potential fragmentation temporarily increases the cross section but has
eventually the opposite effect as microscopic debris is removed by solar
radiation pressure.  In addition, the size of the contribution from increasing
activity of the nucleus itself at \mbox{$r < 3$ AU} is yet to be determined.  On
the other hand, the total sublimation cross section in mid-October seems to have
been at least crudely established to range between 6~km$^2$ and 30~km$^2$.
Interestingly, the size of the largest water-ice depleted pieces of debris
just about equals the size of the debris that, released near 8~AU from the
Sun, was in mid-October, at the Hubble Space Telescope's imaging time, at
the boundary of the inner coma.  Thus, the inner coma, within $\sim$6000~km
or so of the nucleus, still contains sublimating chunks released near 8~AU
from the Sun.
\end{document}